\documentclass[a4paper,12pt]{article}
\usepackage{amsfonts,amsmath,amsthm,amssymb,amscd}
\usepackage{newpxtext,newpxmath} 
\usepackage{natbib}
\usepackage[nobottomtitles*]{titlesec}
\titleformat{\section}{\Large\bf}{\thesection}{.8em}{}
\titlespacing*{\section}{0pt}{*3}{.9ex}[\fill]
 
\usepackage{pict2e} 
\sfcode`P=1000  % normal spacing if P followed by period, as in "LP."
\usepackage{microtype} 
\usepackage{graphicx}
\usepackage[colorlinks,linkcolor=blue,citecolor=blue]{hyperref} 

\usepackage{xcolor}
\newcommand{\supp}{\textup{\small\textsf{supp}}} 
\newcommand{\ix}{\textup{\small\textsf{index}}} 
\newcommand{\sign}{\textup{\small\textsf{sign}}} 

\newcommand{\reals}{\mathbb{R}} 
\newcommand{\eps}{\varepsilon} 
\newcommand{\T}{^{\top}} 
\newcommand{\0}{{\mathbf0}}
\newcommand{\1}{{\mathbf1}}

\def\myproof{\noindent{\textit{Proof.\enspace}}}
\def\endproof{\hfill\strut\nobreak\hfill\tombstone\par\smallbreak}
\def\tombstone{{\large$\square$}}
\newcommand{\maxi}{\mathop{\hbox{\textup{maximize }}}}

\newdimen\einr\einr1.8em
\newdimen\rmeinr\rmeinr1.8em
\newdimen\tmp % intermediate storage 
\newcommand{\abs}[1]{\par\hangafter=1\hangindent=\einr
  \noindent\hbox to\einr{#1\hfill}\ignorespaces} 
\newcommand\rmitem[1]{\abs{\textup{#1}}} 
\newcommand\bullitem{\tmp\einr\einr\rmeinr\abs{\raise.17ex\hbox{\kern7pt\scriptsize$\bullet$}}\einr\tmp}

\parindent\einr
\newtheorem{theorem}{Theorem}
\newtheorem{lemma}{Lemma}
\newtheorem{corollary}{Corollary}
\theoremstyle{definition}
\newtheorem{definition}{Definition}
\newcommand{\sagemath}{{\tt SageMath}}
\newif\ifpictures
\picturestrue
\title
{A Stable-Set Bound and Maximal Numbers of Nash Equilibria in Bimatrix Games}
\author{Constantin Ickstadt\footnote{FB 12 -- Institut f\"ur Mathematik, Goethe-Universit\"at,
  Robert-Mayer-Str.~10, 60325 Frankfurt am Main, Germany.
  Email: ickstadt@math.uni-frankfurt.de,
  theobald@math.uni-frankfurt.de} \and Thorsten Theobald$^*$
  \and Bernhard von Stengel\footnote{Department of
  Mathematics, London School of Economics, London WC2A 2AE,
  United Kingdom. Email: b.von-stengel@lse.ac.uk}}

\date{22 September 2025}
\begin{document}

\maketitle

\begin{abstract}
Quint and Shubik (1997) conjectured that a non-degenerate
$n\times n$ game has at most $2^n-1$ Nash equilibria in
mixed strategies.
The conjecture is true for $n\le 4$ but false for $n\ge6$.
We answer it positively for the remaining case $n=5$, which
had been open since 1999.
The problem can be translated to a combinatorial question
about the vertices of a pair of simple $n$-polytopes with
$2n$ facets.
We introduce a novel obstruction based on the index of an
equilibrium, which states that equilibrium vertices belong
to two equal-sized disjoint stable sets of the graph of the
polytope.
This bound is verified directly using the known
classification of the 159,375 combinatorial types of dual
neighborly polytopes in dimension 5 with 10 facets.
Non-neighborly polytopes are analyzed with additional
combinatorial techniques where the bound is used for their
disjoint facets.

\end{abstract}

% REVISION for MOR
% \title
% {A Stable-Set Bound and Maximal Numbers of Nash Equilibria in Bimatrix Games}
% \author{Constantin Ickstadt\footnote{FB 12 -- Institut f\"ur Mathematik, Goethe-Universit\"at,
%   Robert-Mayer-Str.~10, 60325 Frankfurt am Main, Germany.
%   Email: ickstadt@math.uni-frankfurt.de,
%   theobald@math.uni-frankfurt.de} \and Thorsten Theobald$^*$
%   \and Bernhard von Stengel\footnote{Department of
%   Mathematics, London School of Economics, London WC2A 2AE,
%   United Kingdom. Email: b.von-stengel@lse.ac.uk}}
% 
% % \date{\today}
% \date{19 November 2024}
% \begin{document}
% 
% \maketitle
% 
% \begin{abstract}
% Quint and Shubik (1997) conjectured that a non-degenerate
% $n\times n$ game has at most $2^n-1$ Nash equilibria in
% mixed strategies.
% The conjecture is true for $n\le 4$ but false for $n\ge6$.
% We answer it positively for the remaining case $n=5$, which
% had been open since 1999.
% The problem can be translated to a combinatorial question
% about the vertices of a pair of simple $n$-polytopes with
% $2n$ facets.
% We introduce a novel obstruction based on the index of an
% equilibrium, which states that equilibrium vertices belong
% to two equal-sized disjoint stable sets of the graph of the
% polytope.
% This bound is verified directly using the known
% classification of the 159,375 combinatorial types of dual
% neighborly polytopes in dimension 5 with 10 facets.
% Non-neighborly polytopes are analyzed with additional
% combinatorial techniques where the bound is used for their
% disjoint facets.
% \end{abstract}
% 
\section{Introduction}
% please start each sentence on a new line for easier
% editing; also keeps source text lines short for later
% check for changes with Unix diff command.

A bimatrix game is a two-player game in strategic form, a
basic model of game theory. 
Its central solution concept is \textit{Nash equilibrium},
which is a pair of randomized strategies that are optimal
against each other.
Nash equilibria of bimatrix games are best understood as
certain combinatorial properties of two polytopes derived
from the payoff matrices of the two players, called
\textit{best-response polytopes}.
The possible \textit{number} of Nash equilibria is of
interest for algorithms that find one or all Nash equilibria
of a game. 
Most insights in this area have been found using properties
and constructions of polytopes.
The present paper continues in this tradition with
additional new techniques. 
We answer a long-open problem:
We show that a non-degenerate $5\times 5$ bimatrix game has
at most 31 Nash equilibria
(degenerate games are non-generic and may have infinitely
many Nash equilibria).
Throughout this paper, all the games that we talk about are
assumed to be non-degenerate, which implies that their
best-response polytopes are simple (defined by generic
linear inequalities).

\citet{quint-shubik-1997} conjectured that the maximal
number of Nash equilibria in an $n\times n$ game is $2^n-1$. 
Their evidence for this, not based on polytopes, was that
this is the number of Nash equilibria in the ``coordination
game'' where each player's payoff matrix is the identity
matrix.
The corresponding best-response polytope is the $n$-cube.
For the non-degenerate games that we consider, 
every equilibrium strategy of a player is a separate
polytope vertex, so that the number of vertices is an upper
bound on the number of Nash equilibria.
This implies the Quint-Shubik conjecture for $n\le 3$,
and it was proved for $n=4$ by \citet{keiding-1997} and
\citet{MP-1999}.
However, the conjecture is false for $n\ge6$.
Using dual cyclic polytopes, which for $n\times n$ games
have proportional to about $2.6^n/\sqrt n$ many vertices,
\citet{von-stengel-1999} constructed games with about
$2.4^n/\sqrt n$ many Nash equilibria (slightly fewer for
odd~$n$), and a $6\times 6$ game with 75 Nash equilibria as
the first counterexample to the Quint-Shubik conjecture.
The remaining case $n=5$ had since been open, which we
solve in this paper.

We develop new techniques to obtain our result.
Briefly, they are based on existing lists of combinatorial
types of simple polytopes in dimensions 4 and 5 with 8, 9,
and (for dual-neighborly polytopes) 10 facets, and a novel
\textit{obstruction} that limits the number of equilibrium
vertices in certain face configurations.
A basic version of this obstruction is that a triangle as a
polytope face can have at most two equilibrium vertices,
which ``loses'' one possible equilibrium vertex per triangle.
This basic obstruction has been used by \citet{keiding-1997}
to show that $4\times 4$ games have at most 15 Nash
equilibria, using the list by
\citet{gruenbaum-sreedharan-1967} of the 37 combinatorial
types of simple polytopes in dimension~4 with up to 8 facets
(3~of these 37 polytopes are \textit{dual-neighborly},
that is, any two facets intersect).
For $5\times 5$ games, no such list is known.
We make use of two lists by \citet{firsching-2017}, namely
of the 159,375 dual-neighborly simple polytopes in dimension~5
with 10 facets, and of all 1,142 simple polytopes in
dimension~4 with 9 facets, but need further considerations.

Our novel obstruction is based on the concept of an
equilibrium \textit{index} due to \citet{shapley-1974}.
It is convenient to consider as an equilibrium also the
``artificial equilibrium'', given by the pair of all-zero
vectors that are vertices of the two best-response polytopes
and fulfill the equilibrium condition, but cannot be
re-scaled as randomized strategies to represent a Nash
equilibrium.
The important \textit{parity argument}, based on the
algorithm by \citet{LH} and already exploited by
\citet{keiding-1997}, states that the number of equilibria is
even (and hence the number of Nash equilibria is odd).
The index property strengthens this further:
Half of the equilibria have index~$+1$, and the other half
(including the artificial equilibrium) have index~$-1$.
In a best-response polytope, equilibrium vertices connected
by an edge have opposite index (Lemma~\ref{l:edge} below).
Hence, in the graph of each best-response polytope (as
defined by its vertices and edges), any two index $+1$
equilibrium vertices are non-adjacent and form a
\textit{stable set} (also called independent set),
and the same applies to the index $-1$ equilibrium vertices.
The maximum size of two disjoint stable sets of equal size
is therefore an upper bound on the number of equilibria
(Theorem~\ref{t:stable-sb}), which is our new
\textit{stable-set bound}.
Because the polytope graphs are small, this bound on
equilibrium numbers can be computed quickly.

For dual-neighborly polytopes, the bound of 32 equilibria
(and hence 31 Nash equilibria) for a $5\times 5$ game can
then be verified by computer calculations using the list by
\citet{firsching-2017} of these 159,375 polytopes.
In fact, this result was already obtained in an earlier
unpublished attempt by Vissarion Fisikopoulos and Bernhard
von Stengel using the list of 159,750 neighborly oriented
matroids by \citet{MiyataPadrol2015}, and using the simpler
obstruction of disjoint simplices as faces, each of which
can only contain two equilibrium vertices.
Oriented matroids are more general than polytopes (and hence
the bound is valid), but only slightly so;
\citet{firsching-2017} showed that all but 375 of them are
realizable as polytopes.

The challenge solved in this paper is when a best-response
polytope in a $5\times 5$ game is not dual-neighborly and
therefore has two disjoint facets.
Such a polytope has at most 40 vertices (dual-neighborly
polytopes have 42 vertices).
There is as yet no list of all 5-dimensional simple
polytopes with 10 facets.
The two disjoint facets are 4-dimensional simple polytopes with 
up to 8 facets.
Each of these facets can be seen as a $4\times 4$ game,
which has at most 16 equilibria.
However, this does not complete the proof because there may
be further equilibrium vertices on neither facet.

A crucial observation is that the stable-set bound applies
also to the equilibrium vertices on a \textit{facet} of a
best-response polytope (Theorem~\ref{t:facet}, proved in a
nicely combined use of polytopes and game theory).
Each of the two disjoint facets has one of the 37 combinatorial
types known from \citet{gruenbaum-sreedharan-1967}.
By the new facet-stable-set bound, 35 of these types have at
least 4 non-equilibrium vertices, and the entire polytope
has most 40 vertices.
The remaining two combinatorial types are the 4-cube and what
we call the \textit{semi-cube} (Figure~\ref{fi:semi-cube}).
In order to have more than 32 equilibria, at least one of
the two disjoint facets is therefore a 4-cube or semi-cube;
furthermore, the corresponding facet of the other best-response
polytope needs 9 facets as a polytope in dimension~4.
Again, an exhaustive computer search of these 1,142 simple polytopes,
and the stable-set bound, give combinatorial properties
(Theorem \ref{t:4-9}) that we use to show that more than 32
equilibria are not possible.
This completes the proof that $5\times 5$ games have no
more than 31 Nash equilibria.

For the history of using polyhedra for studying bimatrix
games see \citet{von-stengel-2002}. %more detailed cites later
Cyclic polytopes have been used by \citet{SvS2006, SvS2016}
to show that Lemke-Howson paths (see Section~\ref{s:LH}) may
be exponentially long.
The number of Nash equilibria in $N$-player games has been
studied by \citet{mckelvey-mclennan-1997} and recently
\citet{vujic-2022}.
The more restrictive two-player rank-1 games were introduced by 
\citet{kannan-theobald-2010} and shown to have exponentially
many equilibria by \citet{agm-2021}.
For semi-definite games, equilibrium numbers were
considered by \citet{semidef-games} and \cite{semidef-network-games},
which are relevant for quantum games
(\citet{bostanci-watrous-2021}).
The Quint-Shubik bound does hold for tropical games
(\citet{agm-2023}).

% The paper is structured as follows.

Section~\ref{s:prelim} states preliminaries on bimatrix
games and polytopes.
Best-response polytopes are recalled in Section~\ref{s:brp}.
A concise definition of Lemke-Howson paths and the
equilibrium index for our purposes is given in
Section~\ref{s:LH}.
The stable-set bound is proved in Section~\ref{s:stable}.
Based on these concepts, Section~\ref{s:dim5} gives the main
proof that non-degenerate $5 \times 5$ games have at most
$31$ Nash equilibria.

\section{Preliminaries on bimatrix games and polytopes}
\label{s:prelim}

All matrices have real entries.
The transpose of a matrix $B$ is denoted by $B\T$.
All vectors are column vectors.
We treat vectors and scalars as matrices, so scalars are
multiplied from the right with a column vector and from the
left with a row vector.
The $i$th component of a vector~$x$ is denoted by $x_i$.
The all-zero vector is denoted by~$\0$ and the all-one
vector by~$\1$, their dimension depending on the context.
Let $e_i$ be the $i$-th unit vector.
Inequalities between vectors such as $x\ge\0$ are assumed to
hold for all components.
We let $[k]=\{1,\ldots,k\}$ for any positive integer~$k$.

An $m \times n$ \textit{bimatrix game} is a pair of $m\times
n$ matrices $(A,B)$.
The $m$ rows and $n$ columns are the \textit{pure strategies} of
player 1 and~2, respectively.
The players simultaneously each choose a pure strategy,
with the resulting entry of $A$ as payoff to player~1 and of
$B$ to player~2.
A \textit{mixed strategy} is a probability vector over the
player's pure strategies.
Players are interested in maximizing their expected payoff,
given by $x\T Ay$ to player~1 and $x\T By$ to player~2 if
their mixed strategies are $x$ and~$y$.
A mixed strategy is a \textit{best response} to the other
player's mixed strategy if it maximizes the player's
expected payoff over all possible mixed strategies.
A mixed-strategy pair of mutual best responses is 
called a \emph{Nash equilibrium}.

For a mixed strategy, say $x$ of player~1, its
\textit{support} $\supp(x)$ is the set of pure strategies
that are played with positive probability.
It is easy to see (\citet{Nash1951}, p.~287)
that $x$ is a best response to a mixed
strategy $y$ of player~2
if and only if every pure strategy in $\supp(x)$ is a best
response to~$y$, that is, if the following
\textit{best-response condition} holds:
\begin{equation}
\label{br}
x_i>0\quad\Rightarrow\quad
% (Ay)_i=u=\max\{(Ay)_k\mid 1\le k\le m\} \qquad(1\le i\le m).
(Ay)_i=u=\max_{k\in[m]} \,(Ay)_k \qquad(i\in[m]).
\end{equation}
If (\ref{br}) holds, then $u$ is also the best-response
payoff $x\T Ay$ of the mixed strategy~$x$ against the mixed
strategy~$y$.
The analogous condition holds for a mixed strategy~$y$ of player~2
being a best response to a mixed strategy~$x$ of player~1.

A bimatrix game is \textit{non-degenerate} if for any mixed
strategy $z$ of a player there are at most
$\lvert\supp(z)\rvert$ many pure strategies that are best
responses to~$z$.
For a comprehensive list of equivalent definitions see \citet[thm.~14]{vS21}. 
Generic games $(A,B)$ (where the entries of $A$ and $B$ are
chosen from some continuous distributions) are
non-degenerate.
We only consider non-degenerate games.

We use the following concepts from convex geometry
(see also 
\cite{gruenbaum-2003,ziegler-1995,joswig-theobald-2013}).
An \textit{affine combination} of points $z^1,\ldots, z^k$
% in $\reals^d$ is of the form $\sum_{i=1}^k z^i\lambda_i$
in some Euclidean space is of the form $\sum_{i=1}^k z^i\lambda_i$
where $\lambda_1,\ldots,\lambda_k$ are reals with
${\sum_{i=1}^k \lambda_i=1}$.
It is called a \textit{convex combination} if $\lambda_i\ge0$ for all~$i$.
A set of points is \textit{convex} if it is closed
under forming convex combinations.
The \textit{convex hull} of a set of points is the smallest
convex set that contains all these points.
Given points are \textit{affinely independent} if none of these
points is an affine combination of the others.
A convex set has \textit{dimension}~$d$ if and only if
it has $d+1$, but no more, affinely independent points.

A \textit{polyhedron} $P$ is a subset of $\reals^d$ defined
by finitely many linear inequalities, that is,
$P=\{z\in\reals^d\mid Cz\le q\}$ for some $C$ in
$\reals^{\ell\times d}$ and some $q$~in~$\reals^\ell$.
A \textit{face} $F$ of the polyhedron $P$ is obtained by
converting some of its inequalities into equalities, that
is, if for some $S\subseteq \{1,\ldots,\ell\}$
\begin{equation}
\label{face}
F=\{z\in\reals^d\mid Cz\le q,~ (Cz)_i=q_i \hbox{ for }i\in
S\}\,.
\end{equation}
A \textit{polytope} is a bounded polyhedron.
A \textit{simplex} is the convex hull of affinely
independent points.

Suppose the polytope $P$ has dimension $d$
(also called a $d$-polytope;
as a subset of $\reals^d$ it is then called \textit{full-dimensional}).
Then a face $F$ of $P$ is called
a \textit{facet} if it has dimension $d-1$, 
a \textit{ridge} if it has dimension $d-2$, 
an \textit{edge} if it has dimension $1$, 
and a \textit{vertex} if it has dimension~$0$. 
A vertex is a singleton $\{v\}$ and also identified
with~$v$. 
% A \textit{$k$-face} is a face of $P$ of dimension~$k$.

A $d$-polytope $P$ is called \textit{simple} if
every vertex of $P$ belongs to exactly $d$ facets.
For the following more general statement see \citet[prop.~2.16(iv)]{ziegler-1995}.

\begin{lemma}
\label{l:kface}
% In a simple $d$-dimensional polytope $P$, every non-empty
In a simple $d$-polytope $P$, every non-empty
face of $P$ of dimension $k$ is a subset of exactly $d-k$
facets.
\end{lemma}

The faces of a polytope, partially ordered by inclusion,
form a lattice (see \cite{ziegler-1995}).
Two polytopes have the same \emph{combinatorial type}
if their face lattices are isomorphic.
We denote the set of combinatorial types of $d$-dimensional
simple polytopes with $k$ facets by $\mathcal{P}_{k}^d$.

A simple $d$-polytope is \textit{dual-neighborly} if any 
$\lfloor d/2\rfloor$ of its facets have a non-empty
intersection.
The dimension $d$ of the polytopes that we consider is $4$
or~$5$, so they are dual-neighorly if any two facets have a
non-empty intersection, which is then a ridge by
Lemma~\ref{l:kface}.

\begin{lemma}
\label{l:neighb}
For a given dimension and number of facets, only simple
polytopes that are dual-neighborly have the largest numbers
of vertices.
The maximal number of vertices of a polytope in
$\mathcal{P}_{8}^4$, $\mathcal{P}_{9}^4$, $\mathcal{P}_{9}^5$,
$\mathcal{P}_{10}^5$
is 20, 27, 30, 42, respectively.
\end{lemma}

\myproof
See \citet[p.~174]{gruenbaum-2003}
and (3.2) in \citet{joswig-theobald-2013}.
\endproof

The \textit{graph} of a polytope is defined by the vertices
and edges of the polytope, where in the graph an edge is
considered as the unordered pair of its endpoints, written
as $uv$ if the endpoints are $u$ and~$v$; then $u$ and $v$
are also called \textit{adjacent}.
The \textit{degree} of a vertex in a graph is the number of
vertices it is adjacent~to.

\section{Best-response polytopes}
\label{s:brp}

For more details on the following construction of
``best-response polytopes'' see
\citet{von-stengel-1999,von-stengel-2002,vS2022}. 
Let $(A,B)$ be an $m\times n$ bimatrix game and consider the
polyhedra
\begin{equation}
\label{PQ}
\arraycolsep.2em
\begin{array}{rcllrclrcl}
P &=&  \{~x \in \reals^m & \mid & x&\ge& \0, & B\T x&\le&\1~\},\\
Q &=&  \{~y \in \reals^n & \mid & A y& \le& \1, & y& \ge&\0~\}.
\end{array}
\end{equation}
We assume that $P$ and $Q$ are polytopes, which holds if and
only if the best-response payoff to any mixed strategy of
the other player is always positive \citep[lemma~9.9]{vS2022}.
A sufficient %(but not necessary)
condition for this is that $A$ and $B\T $ are nonnegative
and have no zero column.
This is not restrictive since adding a constant to every
entry of a payoff matrix does not change best responses.

With the exception of $\0$, every $x$ in $P$ represents a
mixed strategy of player~1 after re-scaling it to
$x\frac1{\1\T x}$ as a probability distribution.
Similarly, every $y$ in $Q\setminus\{\0\}$ represents
a mixed strategy $y\frac1{\1\T y}$ of player~2.
We always consider $x$ in $P\setminus\{\0\}$ and $y$ in
$Q\setminus\{\0\}$ as ``mixed strategies'' with this implied
re-scaling.

Suppose that $j\in[n]$ and consider the last $n$
inequalities $B\T x\le\1$ in the definition of~$P$ in
(\ref{PQ}).
If the $j$th inequality is \textit{binding} (holds as
equality), that is, $(B\T x)_j=1$, then $j$ is a pure best
response of player~2 to~$x$.
That is, the right-hand side~$1$ in every inequality in
$B\T x\le\1$ represents the best-response payoff to~$x$,
re-scaled as~$1$, if at least one of these inequalities is
binding; the actual payoff is the re-scaling factor
$\frac1{\1\T x}$.
Similarly, a binding inequality $(Ay)_i=1$ for $i\in[m]$
among the inequalities $Ay\le\1$ in the definition of~$Q$
means that $i$ is a pure best response to~$y$.

If $x=\0$ then $B\T x<\1$ and if $y=\0$ then $Ay<\1$.
Hence, $P$ and $Q$ contain the affinely independent
vectors $\0$ and $e_i\eps$ for the unit vectors~$e_i$
(in $\reals^m$ respectively $\reals^n$)
for sufficiently small
positive~$\eps$ and are full-dimensional.

We introduce \textit{labels} in $[m+n]$ to uniquely identify 
the pure strategies of player~1 as 
$1,\ldots,m$ and of player~2 as $m+1,\ldots,m+n$.
The inequalities in (\ref{PQ}) are written in the order of
these labels.
For a point $x\in P$ or~$y\in Q$, \textit{its labels} are the
labels of the \textit{binding} inequalities for that point.
That is, for a pure strategy $i$ in $[m]$ of player~1,
the point $x$ has label~$i$ if $x_i=0$, and $y$ has
label~$i$ if $i$ is a best response to the 
mixed strategy~$y$ (if $y=\0$ then $Ay<\1$ and
$y$ cannot have label~$i$).
Similarly, for a pure strategy $j$ in $[n]$ of player~2, the
mixed strategy~$x$ has label~$m+j$ if $j$ is a best response
to~$x$, and $y$ has label~$m+j$ if $y_j=0$.
We also consider these labels if
$x=\0$ (then $x$ has labels $1,\ldots,m$)
or $y=\0$ (then $y$ has labels $m+1,\ldots,m+n$).

By the best-response condition~(\ref{br}), $(x,y)$ is a Nash
equilibrium of $(A,B)$ if and only if every pure strategy of
a player is a best response or played with probability zero
(or both).
That is, every label in $[m+n]$ has to appear as a label of
$x$ or~$y$.

\begin{definition}
Consider a bimatrix game $(A,B)$ and assume that $P$ and $Q$
in $(\ref{PQ})$ are polytopes, called the
\emph{best-response polytopes} for the game.
Let $(x,y)\in P\times Q$ be such that every element of
$[m+n]$ appears as a label of $x$ or~$y$.
Then $(x,y)$ is called \emph{completely labeled} and an
\emph{equilibrium} of the game.
This includes the \emph{artificial equilibrium} $(\0,\0)$.
Any other equilibrium, with $x$ and $y$ re-scaled as mixed
strategies, is a Nash equilibrium of $(A,B)$.
\end{definition}

Note that if $x\in P\setminus\{\0\}$, then $x$ is missing at
least one label $i\in[m]$, which in an equilibrium $(x,y)$
has to appear as a label of $y$ and therefore $y\ne\0$.
Hence $(\0,\0)$ is the only equilibrium $(x,y)$ where $x=\0$ or
$y=\0$.

The Nash equilibria $(x,y)$ of a bimatrix game are exactly the
completely labeled pairs of points in $P\times Q\setminus
\{(\0,\0)\}$, even for a degenerate game.
For a non-degenerate game, they are always pairs of vertices.

\begin{lemma}
\label{l:nondeg}
For a bimatrix game $(A,B)$ with best-response polytopes $P$
and $Q$ , the following are equivalent:
\rmitem{(a)}
The game is non-degenerate.
\rmitem{(b)}
No point in $P$ has more than $m$ labels 
and no point in $Q$ has more than $n$ labels.
\rmitem{(c)}
$P$ and $Q$ are simple polytopes, and for both polytopes any
redundant inequality (which can be omitted
without changing the polytope) is never binding.

\noindent
Furthermore, assume the game is non-degenerate and let
$(x,y)\in P\times Q$ be an equilibrium. 
\rmitem{(d)}
Then $x$ is a vertex of $P$ and $y$ is a vertex $Q$, and $x$
has exactly $m$ labels and $y$ has exactly the other $n$ labels. 
\end{lemma}

\myproof
Conditions (b) and (c) are theorem~14 (b) and (h) of
% \cite[thm.]{vS21}. 
\cite{vS21}. 
Because there are only $m+n$ labels in total, (b)
implies~(d).
\endproof

In Lemma~\ref{l:nondeg}(c), redundant inequalities refer to
the description of the polytope, which may not affect
whether the polytope as a set is simple.
It implies, for example, that there are no duplicate
inequalities.
A never-binding inequality represents a pure strategy that
is never a best response; this does not affect
non-degeneracy and may exist in a generic game, but we can
omit such a strategy from the game.
In the non-degenerate games that we consider, every binding
inequality defines a facet that has its own label.

The transition from bimatrix games to polytopes can also be
reversed. 
For projective transformations see \citet{ziegler-1995}, of
which (\ref{proj}) is a simple form.

\begin{lemma}
\label{l:biject}
Consider a simple $m$-polytope $P$ and a simple $n$-polytope $Q$,
each with $m+n$ facets labeled $1,\ldots,m+n$.
Let $(x_0,y_0)$ be a completely labeled vertex pair of
$P\times Q$.
Then there is a game $(A',B')$ with best-response polytopes
$P'$ and $Q'$ such that:
\rmitem{(a)}
There are affine bijections $P\to P'$ and $Q\to Q'$
with a corresponding permutation of $[m+n]$ that maps the
facet labels of $P$ and $Q$ to those of $P'$ and $Q'$,
with $(x_0,y_0)$ mapped to $(\0,\0)$ in $P'\times Q'$,
such that all completely labeled vertex pairs of $P\times Q$
are in bijection to those of $P'\times Q'$.
\rmitem{(b)}
By applying, if necessary, bijective projective
transformations to the polytopes $P'$ and $Q'$ that leave
their face lattices unchanged, all entries of $A'$ and
$B'$ are positive.
\end{lemma}

\myproof
For (a) see \citet[prop.~2.1]{von-stengel-1999}.
For (b), note that e.g.\ $Q'$, given by $\{y'\in\reals^n\mid
{A'y'\le\1},~{y'\ge\0\}}$, may be a polytope even when $A'$ has
non-positive entries.
In that case add a sufficiently positive constant $\alpha$
to all entries of $A'$ to yield an all-positive matrix $\hat
A=A'+\1\alpha\1\T$, and let
$\hat Q= \{\hat y\in\reals^n\mid \hat A\hat y\le\1,~\hat y\ge\0\}$.
We claim that
\begin{equation}
\label{proj}
y'\mapsto \hat y = y'\frac{1}{1+\alpha\,\1\T y'}
\end{equation} 
is a bijective projective transformation $Q'\to\hat Q$ that
preserves all binding inequalities and therefore the face
lattices of $Q'$ and~$\hat Q$.
In (\ref{proj}), $\0$ in $Q'$ is mapped to $\0$ in $\hat Q$,
and for $y'\ne\0$ the mapping in (\ref{proj}) is obtained by
mapping $y'$ to $(\overline y,u)$ with a mixed strategy
$\overline y=y'u$ where $u=\frac1{\1\T y'}$, then
$(\overline y,u)$ to $(\overline y,u+\alpha)$, and this is
mapped to $\hat y=\overline y\frac1{u+\alpha}$.
To see that binding inequalities are preserved % correspondingly
in $Q'$ and~$\hat Q$, clearly
$y'_j=0$ if and only if $\hat y_j=0$ for $j\in[n]$.
Furthermore, $\hat A\hat y\le\1$ if and only if
$\hat A y'\le \1(1+\alpha\1\T y')$, which for the $i$th row
$a_i\T$ of $A'$ means
$
(a_i\T + \alpha \1\T)y'\le 1+\alpha\1\T y'
$
or equivalently $a_i\T y'\le 1$, which is the $i$th
inequality in $A'y'\le\1$. 

A similar transformation changes $B'$ to an all-positive
matrix $\hat B$ and $P'$ to a ``projectively equivalent''
polytope~$\hat P$.
\endproof

Lemma~\ref{l:biject} allows changing the labels of $P$
and $Q$ such that a Nash equilibrium and the artificial
equilibrium exchange roles, which we will make use of in
Theorem~\ref{t:facet}.

\section{Equilibria come in pairs of opposite index}
\label{s:LH}

Our ``stable-set bound'' is based on the fact that
equilibria come in pairs of opposite \textit{index},
which we explain in this section.
The algorithm by \citet{LH} starts at one equilibrium by
following a path of alternating edges in the best-response
polytopes $P$ and $Q$ in (\ref{PQ}) where one of the labels
is allowed to be \textit{missing}, until it terminates at
another equilibrium.
The endpoints of the Lemke-Howson paths are equilibria of 
opposite index. 
The most concise description, given here, considers the
product polytope $Z=P\times Q$.
For more details %with examples and pictures
see, for example, \citet{von-stengel-2002,vS21,vS2022}.

\begin{definition}
\label{d:labpol}
A \emph{labeled polytope} $Z$ is a simple $d$-polytope of
the form 
\begin{equation}
\label{labpol}
Z =\{z\in \reals^d\mid -z\le\0,~ Cz\le \1 \}
\end{equation}
for some $k\times d$ matrix $C$ without redundant
inequalities where each binding inequality has a label in $[d]$,
such that:
For $i\in[d]$, the $i$th inequality $-z_i\le0$ in $-z\le\0$ has
label~$i$, and for $j\in[k]$, the $j$th inequality
$(Cz)_j\le 1$ in $Cz\le\1$ has some given label
$\ell(j)$ in $[d]$. 
A point $z$ in $Z$ that has all labels in $[d]$ is called
\textit{completely labeled}, and
\textit{$h$-almost completely labeled} if it has all labels
in $[d]\setminus\{h\}$, for some $h\in[d]$.
\end{definition}

We will apply this to $Z=P\times Q$ where in Definition~\ref{d:labpol}
we take $d=m+n$, $z=(x,y)$, and
\begin{equation}
\label{CAB}
C=\left[\begin{matrix}0&A\\B\T & 0\end{matrix}\right],
\quad k=d, \quad\hbox{and}\quad
\ell(j)=j ~~\hbox{ for }j\in[k]\,.
\end{equation}

\begin{theorem}
\label{t:even}
Any labeled polytope has an even number of completely
labeled points.
\end{theorem} 

\myproof
Let $Z$ be the labeled polytope as in Definition~\ref{d:labpol}.
Because no inequality is redundant, every facet is defined
by a binding inequality and has a unique label.
Every vertex of $Z$ belongs to exactly $d$ facets and is
incident to exactly $d$ edges, obtained by allowing one of
the binding inequalities to be non-binding (visualized as
``moving away'' from the respective facet).

Consider a fixed label $h$ in $[d]$ that is allowed to be
``missing'', and all the points in $Z$ that are
$h$-almost completely labeled.
Those points that have each label in $[d]\setminus\{h\}$
exactly once define edges of $Z$ except for the endpoints of
these edges.
Their endpoints, which are vertices of $Z$,
are either completely labeled, or are missing label $h$ and
therefore have a \textit{duplicate} label.
Any such vertex with the duplicate label belongs to two
facets that have that label, and is incident to \textit{two}
$h$-almost completely labeled edges, which belong to only
one of the two facets.

Hence, the $h$-almost completely labeled points of $Z$
are vertices and edges of $Z$ that define a subgraph of the
graph of $Z$ where every vertex has either degree two (with
a duplicate label but missing label~$h$) or degree one (if
the vertex is completely labeled).
Such a graph consists of paths and cycles, and every path
has two endpoints, which are the completely labeled
vertices, so their number is even. 
\endproof

Completely labeled vertices (except $\0$) of a labeled
polytope also represent Nash equilibria of a game, namely
of the $d\times k$ game $(U,C\T)$ where
$U=[e_{\ell(1)}\cdots e_{\ell(k)}]$
is composed of the unit vectors for the labels of the rows
of~$C$.
This simplifies certain constructions of games by using only a
single labeled polytope, see \citet{SvS2016}.
Applied to $C$ in (\ref{CAB}), $U$ is the $d\times d$
identity matrix, whose Nash equilibria $(z,w)$ correspond to the
\textit{symmetric} Nash equilibria $(z,z)$ of the symmetric
game $(C,C\T)$, which we seek when setting $z=(x,y)$ to find
the Nash equilibria $(x,y)$ of $(A,B)$.

\begin{definition}
\label{d:index}
Consider a labeled $d$-polytope $Z$ as in
Definition~\ref{d:labpol} and a completely labeled vertex
$z$ with its $d$ binding inequalities written as
$a_i\T z=\beta_i$ for each label $i\in[d]$.
That is, if the $i$th inequality in $-z\le\0$ is binding
then $a_i\T z=\beta_i$ stands for $-z_i=0$ with $a_i$ as the
negative $-e_i$ of the $i$th unit vector~$e_i$, 
and if $i=\ell(j)$ then the $j$th inequality in $Cz\le\1$
has label~$i$ and $a_i\T z=\beta_i$ stands for $(Cz)_j=1$ 
with $a_i\T$ as the $j$th row of~$C$.
Writing the facet normal vectors $a_1,\ldots,a_d$ in the order of their
labels as a matrix with determinant $|a_1\cdots a_d|$,
we define the \emph{index} %$\ix(z)$
of $z$ as the sign of this determinant
(times $-1$ if $d$ is even),
\begin{equation}
\label{index}
\ix(z)=(-1)^{d+1}\sign(|a_1\cdots a_d|).
\end{equation}
\end{definition}

Because the polytope $Z$ is simple, the normal vectors $a_i$
of the incident facets of a vertex are linearly independent
and the determinant is non-zero.
In (\ref{index}), the minus sign for even $d$ has the
purpose of ensuring that the artificial equilibrium~$\0$ has
index $-1$, because the determinant of the negative of the
$d\times d$ identity matrix is $(-1)^d$.

The following theorem is due to \citet{shapley-1974}.
For accessibility in the present context, we outline a
streamlined version of its proof due to \citet[thm.~13]{vS21}.

\begin{theorem}
\label{t:index}
For a labeled polytope, the endpoints of any $h$-almost
completely labeled path are completely labeled vertices of
opposite index.
Hence, half of the completely labeled vertices have index $+1$
and the other half, including $\0$, have index $-1$. 
\end{theorem}

\myproof
The proof extends that of Theorem~\ref{t:even} by
\textit{orienting} each $h$-almost completely
labeled edge to give a consistent orientation of the
resulting paths.
The orientation of the edge is from its endpoint with
negative index to the endpoint with positive index, where
the definition of index is extended to $h$-almost completely
labeled vertices as follows.

Suppose for simplicity that $h=1$ and that $d$ is odd so
that the signs of index and determinant coincide (for even
$d$ they are opposite and the same argument applies).
Consider a completely labeled vertex $z$ of negative index
with normal facet vectors $a_1,\ldots,a_d$ in the order of
their labels, that is, $|a_1\cdots a_d|<0$.
Let the $1$-almost completely labeled edge connect $z$ to $w$ where 
the facet normal vector $a_1$ is replaced by $b_g$ with
label $g$ in~$[d]$.
Then $|b_g\,a_2\cdots a_d|>0$, because the determinant of
the endpoints of any edge of a simple polytope have opposite
sign when replacing the changed facet normal vector and
leaving the unchanged facet normal vectors in place
\citep[lemma~12]{vS21}.

If $g$ is the missing label~$1$ then we are done, that is,
$w$ is the endpoint of the path and has positive determinant.
Otherwise, label $g$ is duplicate.
We now exchange $b_g$ with $a_g$ in writing down the determinant, which 
changes sign, that is, $|a_g\,a_2\cdots b_g \cdots a_d|<0$. 
Then $a_g$ is again the normal vector of the facet 
that the next edge on the path moves away from, and we
proceed as before.
In that way, all the $h$-almost completely labeled edges 
are oriented in the same direction on the path. 
\endproof

\section{The stable-set bound}
\label{s:stable}

In the previous section, we have considered a general labeled
polytope~$Z$ to explain equilibria as endpoints of oriented
paths on $Z$ defined by the points of $Z$ that have all
labels except a missing label~$h$.
For bounds on the number of equilibria of an $m\times n$
game $(A,B)$, it is better to consider the product structure
$Z=P\times Q$ with the $m$-polytope $P$ and $n$-polytope $Q$
in~(\ref{PQ}), because they have much fewer vertices than a
general simple polytope $Z$ of dimension $m+n$.
We assume throughout that $P$ and $Q$ are the simple
polytopes in (\ref{PQ}), with their $m+n$ inequalities
having the respective labels in $[m+n]$ when they are
binding and thus defining a facet of the polytope.

By Lemma~\ref{l:nondeg}(d), any equilibrium is a vertex pair
$(x,y)$ of $P\times Q$. 
We then call $x$ an \textit{equilibrium vertex} of $P$
and the vertex $y$ of $Q$ its \textit{partner} (and vice
versa).
The partner $y$ of $x$ is \textit{unique} because it is
defined by the unique set of labels in $[m+n]$ (and the
corresponding facets of~$Q$) that are missing from~$x$, by
Lemma~\ref{l:nondeg}(d).

The index of an equilibrium $(x,y)$ cannot be told from the
equilibrium vertex~$x$ and its position in~$P$ alone
(other than by uniquely identifying its partner~$y$).
However, the following lemma applies.

\begin{lemma}
\label{l:edge}
Any two equilibrium vertices of $P$ % or $Q$
that are connected by an edge belong to equilibria of opposite index.
\end{lemma} 

\myproof 
Let $x$ and $x'$ be two equilibrium vertices of $P$ with
partners $y$ and $y'$ in $Q$, respectively, and let $xx'$ be
an edge in the graph of~$P$.
Then $x$ and $x'$ differ by exactly one label, where the
set of labels of~$x$ is $\{h\}\cup K$ and of $x'$ is
$\{g\}\cup K$, with $|K|=m-1$ and $h\ne g$.
Because $(x,y)$ and $(x',y')$ are completely labeled,
the set of labels of $y$ is therefore 
$\{g\}\cup L$ and of $y'$ is $\{h\}\cup L$, with
$L=[m+n]\setminus(K\cup\{h,g\})$.
Hence, $yy'$ is also an edge of the graph of~$Q$.
In the graph of the polytope $P\times Q$ (which is the
``product graph'' of the graphs of $P$ and $Q$),
the two equilibria are therefore the endpoints of the
$h$-almost completely labeled path given by the sequence of
vertex pairs $(x,y)-(x',y)-(x',y')$ and therefore have
opposite index by Theorem~\ref{t:index}.
\endproof 

For any undirected graph $G$, a \textit{stable set}
is a set of vertices no two of which are adjacent.

\begin{theorem}[Stable-set bound]
\label{t:stable-sb}
Let $G$ be the graph of $P$ and $S_1$ and $S_2$ be two
disjoint stable sets of~$G$ of equal size $|S_1|$, and let this
size be maximal among all such pairs.
Then the game has at most $2|S_1|$ equilibria.
\end{theorem}

\myproof 
By Theorem~\ref{t:index}, half of all equilibria have
positive index.
No two of their equilibrium vertices in $P$ are adjacent in
$G$ by Lemma~\ref{l:edge}, so they form a stable set $S$
of~$G$.
The same applies to the equilibria of negative index, whose
equilibrium vertices in $P$ form a stable set of~$G$
disjoint from $S$ of the same size.
\endproof 

A \textit{clique} of a graph is a set of vertices every two
of which are adjacent.
Disjoint cliques of size three or larger provide a bound
that is weaker than the stable-set bound, but may match it
(in particular when reduced by~1 to obtain an even number
using Theorem~\ref{t:even}).

\begin{corollary}[Disjoint-clique bound]
\label{c:clique}
Suppose the graph of $P$ has $V$ vertices and $k$ pairwise
disjoint cliques of sizes $c_1,\ldots,c_k$\,.
Then every clique contains at most two equilibrium vertices,
that is, the game has at most $V-\sum_{i=1}^k (c_i-2)$ many equilibria.
This bound is at least as large as the stable-set bound.
\end{corollary}

\myproof 
As in the proof of Theorem~\ref{t:stable-sb}, no two
equilibrium vertices for equilibria of the same index can
belong to a clique because they would be adjacent, in
contradiction to Lemma~\ref{l:edge}.
\endproof 

Corollary~\ref{c:clique} has been known and used earlier
by \citet{keiding-1997} for $4\times 4$ games and triangles
as cliques.
It can be proved without the concept of an equilibrium
index, as follows, here for general $m\times n$ games.

\noindent\textit{Alternative Proof of Corollary~\ref{c:clique}.}\enspace
Consider a clique of the graph of $P$ of size $c$.
The convex hull $F$ of its vertices is a simplex of
dimension $c-1$ and therefore the intersection of
$m-c+1$ facets of~$P$ with as many labels, by Lemma~\ref{l:kface}.
Considered as a $(c-1)$-polytope, $F$ has $c$ facets, each
of which is obtained by intersecting $F$ with an additional
facet of~$P$ that provides an additional label.
The labels of all vertices of $F$ belong therefore to a set
of size $m+1$.
Of these, any equilibrium vertex needs a partner in $Q$
that has the missing $m+n-(m+1)$, that is, $n-1$ labels,
all of which belong therefore to a face of $Q$ of
dimension~1, which is an edge.
An edge has only two endpoints, so at most two vertices in
the clique can be equilibrium vertices.
\endproof

The stable-set bound in Theorem~\ref{t:stable-sb} is our
main tool for proving that $5\times 5$ games have at most 32
equilibria.
It is enhanced by the corresponding property for the
equilibrium vertices on a facet of~$P$.

\begin{theorem}[Facet-stable-set bound]
\label{t:facet}
Let $m\ge2$ and consider a non-degenerate $m\times n$ game $(A,B)$ with
positive payoff matrices and best-response polytopes $P$
and~$Q$.
Consider a facet $F$ with label $\ell$ of $P$
and the graph $G_F$ of $F$.
Then the equilibrium vertices in $G_F$ are contained in two
disjoint stable sets in $G_F$ of equal size.
\end{theorem} 

\myproof 
We can assume that $F$ has at least one equilibrium
vertex~$x_0$ with partner $y_0$ in $Q$.
By assumption, $x_0$ has label~$\ell$.
If $\ell\not\in[m]$ then $\ell=m+j$ for some pure strategy
$j\in[n]$ of player~2.
Then we use the bijections in Lemma~\ref{l:biject} to
map $(x_0,y_0)$ to a new polytope pair $P'\times Q'$ such
that the Nash equilibrium $(x_0,y_0)$ of $(A,B)$ maps to the
artificial equilibrium $(\0,\0)$ of a new game $(A',B')$ and
therefore $\ell$ is in bijection to a strategy of player~1.
Furthermore, $A'$ and $B'$ have positive entries by
Lemma~\ref{l:biject}(b).
We then consider $(A',B')$ instead of $(A,B)$, with the same
number of equilibria for the two games, and $\ell$
(replaced by its image under the bijection) now denoting a
strategy of player~1. 
The polytopes $P$ and $Q$ and facet $F$ of $P$ are also
replaced accordingly.

In this (if needed, new) game, consider all the equilibria
$(x,y)$ such that $x$ is a vertex of $G_F$.
Because $x\in F$, these equilibria
(including the artificial equilibrium $(\0,\0)$)
have label $\ell$ in $[m]$ with $x_\ell=0$, that is,
row~$\ell$ is not played.
All the Nash equilibria among them are also Nash equilibria
of the smaller $(m-1)\times n$ game obtained from $(A,B)$ by
deleting row~$\ell$ from both $A$ and~$B$.
(The smaller game may have additional Nash equilibria.)
The smaller game has $F$ as its best-response polytope for
player~1, and $Q$ with inequality $\ell$ omitted as the
best-response polytope~$Q'$ for player~2.

The facet $F$ is simple because $P$ is simple.
The polyhedron $Q'$ is a polytope because all entries of $A$
are positive.
However, $Q'$ may not be simple and then the smaller game
may be degenerate. 
However, $F$ has label~$\ell$ and therefore the partners of
the equilibrium vertices in $F$ are not affected by omitting
the inequality with label $\ell$ in $Q$ and are the same
vertices in~$Q'$.
(Alternatively, one could perturb $A$ without
changing the combinatorial structure of~$Q$ to create a
simple polytope~$Q'$.)

The edges in $G_F$ are exactly the edges of the graph of~$P$
with endpoints in~$F$.
Theorem~\ref{t:stable-sb} applies to the smaller game with $F$
instead of $P$, which proves the claim. 
\endproof 

Theorem~\ref{t:facet} is not trivial because it is
conceivable that there are, say, $k+1$ equilibrium vertices in
$F$ belonging to equilibria of $(A,B)$ of positive index, and
$k$ for negative index, but with a facet-stable-set bound of~$2k$.
The theorem precludes this possibility. 

\section{Five-by-five games}
\label{s:dim5}

We show that the Quint-Shubik conjecture holds for $5 \times 5$ games.

\begin{theorem}
\label{t:5x5}
Any non-degenerate $5 \times 5$ game has at most 31 Nash equilibria.
\end{theorem}

The proof uses some case distinctions and relies on computer calculations.

In the following, we let $m=n=5$ and consider the two
best-response polytopes $P$ and $Q$ in (\ref{PQ}), both of
which are full-dimensional as shown in Section~\ref{s:brp}
and therefore of dimension~5. They are simple
by Lemma~\ref{l:nondeg}.
By their definition in~\eqref{PQ}, each of
$P$ and $Q$ has at most $5+5=10$ facets.
We can assume that $P$ and $Q$ have ten facets each.
If, say, $P$ has at most nine facets, then
$P$ has at most 30 vertices by Lemma~\ref{l:neighb}
and Theorem~\ref{t:5x5} holds immediately.

We first deal with the case that at least one of the polytopes 
$P$ or $Q$ is dual-neighborly.
Recall that we also consider the artificial equilibrium
$(\0,\0)$ in $P\times Q$ as an equilibrium.

\begin{theorem}
\label{t:neighb}
Consider the best-response polytopes $P$ and $Q$ for a
$5\times 5$ game.
If $P$ or $Q$ is dual-neighborly, then the game has at most
$32$ equilibria.
\end{theorem}

\myproof
Assume that $P$ is dual-neighborly. 
The 159,375 combinatorial types of dual-neighborly simple
5-polytopes with 10 facets have been classified by
\citet{firsching-2017}.
For each such combinatorial type we have verified
the stable-set bound of Theorem~\ref{t:stable-sb} as not
exceeding~32 by computer calculations. 
Indeed, for each of those polytopes, the size of a largest
stable set is at most 16, which implies that the stable-set
bound is at most~32.
The calculations were carried out in {\sagemath} using the
coordinate descriptions of the polytopes from the data in
\cite{firsching-2017}.
We have documented the data of our stable set computations
with links to the respective Sage programs and input files
in~\citet{stable-data}.
\endproof

As mentioned in the introduction, Theorem~\ref{t:neighb}
was already found earlier with the help of Vissarion
Fisikopoulos using the list of 159,750 neighborly oriented
matroids by \citet{MiyataPadrol2015} with the
disjoint-clique bound of Corollary~\ref{c:clique}.
% \fbox{cite as unpublished paper?}

 From now on we assume that $P$ and $Q$ are
\textit{non-neighborly}, that is, not dual-neighborly.
Currently, no list of all polytopes (not only the dual-neighborly
ones) in $\mathcal{P}_{10}^5$ is available.
Even if that list existed, it is unclear if finding the
stable-set bounds would be computationally feasible and if
they would produce the desired bound.

Instead we study the combinatorial structure of non-neighborly 
polytopes $P$ and $Q$.
Because simple 5-polytopes with 10 facets are dual-neighborly
if and only if they have $42$ vertices 
(by Lemma~\ref{l:neighb}), $P$ and $Q$ have at
most 41 vertices.
In fact, as simple 5-polytopes, $P$ and $Q$ have an even
number of vertices, because every vertex in their graph
has degree~5, and the number of odd-degree vertices in a
graph is even (because the sum of degrees is twice the
number of edges).
The interesting numbers of vertices of $P$ and $Q$ for
proving Theorem~\ref{t:5x5} are therefore 40, 38, 36, and~34.
Because the number of equilibria is even by
Theorem~\ref{t:even}, it suffices to know that seven vertices in
$P$ cannot be equilibrium vertices.
We call such vertices \textit{obstruction vertices}.

Because $P$ is non-neighborly, $P$ has two facets that are
disjoint.
We study the facet-stable-set bound stated in
Theorem~\ref{t:facet} for these two facets.
For the dual-neighborly 5-polytopes in Theorem~\ref{t:neighb},
it was sufficient to find maximal single stable sets % $S_1$
to bound the equilibrium numbers. % with $2|S_1|$.
For the four-dimensional facets, we need to consider pairs
of disjoint stable sets as in Theorem~\ref{t:stable-sb}.

\begin{lemma}
\label{l:ilp}
The stable-set bound for pairs of disjoint stable sets
is the optimal value of an integer linear program (ILP)
in $2V$ binary variables if $P$ has $V$ vertices.
\end{lemma}

\myproof
Let $G$ be the graph of the polytope $P$ with, for
simplicity, vertex set $[V]$ and edge set $E$.
The maximum size of a stable set in $G$
can be computed by the integer linear program
\[
  \begin{array}{rcl}
  \multicolumn{3}{l}{\maxi \sum_{i=1}^V x_i} \\
  \text{subject to }\;\; x_i + x_j & \le & 1 \quad \text{ for all } ij \in E \, , \\
  x & \in & \{0,1\}^V \, ,
  \end{array}
\]
where the vector $x$ is the incidence vector of a stable set.
In order to compute the stable set bound, we consider the modified
ILP in $2V$ variables
\[
  \begin{array}{rcl}
  \multicolumn{3}{l}{\maxi \sum_{i=1}^V x_i + \sum_{i=1}^V y_i} \\
  \text{subject to }\;\; x_i + x_j & \le & 1 \quad \text{ for all } ij \in E \, , \\
  y_i + y_j & \le & 1 \quad \text{ for all } ij \in E \, , \\
  x_i + y_i & \le & 1 \quad \text{ for all } i \in [V] \, , \\
  \sum_{i=1}^V x_i & = & \sum_{i=1}^V y_i \, , \\  
  x,y & \in & \{0,1\}^V \, .
  \end{array}
\]
Both $x$ and $y$ provide incidence vectors of stable sets. The conditions
$x_i + y_i \le 1$ enforce the disjointness of these stable sets and
the condition $\sum_{i=1}^V x_i = \sum_{i=1}^V y_i$ ensures
that the stable sets specified by $x$ and $y$ have the same size.
Then the objective function gives the sum of the sizes of the two
stable sets, or, equivalently, twice of size of either of them.
\endproof

As in the proof of Theorem~\ref{t:facet}, it does not
matter if the labels of the two disjoint facets of $P$ denote pure
strategies of player~1 or player~2 or of both players.
Hence, we assume that the two disjoint facets of $P$ have
labels 1 and~2 (representing the first two strategies of
player~1, but this is irrelevant), and call the facets $P_1$
and~$P_2$.
Similarly, let $Q_1$ and $Q_2$ be the (not necessarily
disjoint) facets of $Q$ with labels 1 and~2, respectively.

Each facet $P_1$ and $P_2$ of $P$ is a four-dimensional
polytope with at most eight facets.
Every other facet of $P$, as a 4-polytope, may have up to nine
(three-dimensional) facets, obtained as intersections with
the other nine facets of~$P$.

A computer calculation of the facet-stable-set bound of
Theorem~\ref{t:facet} shows that 35 out of the 37 combinatorial
types of polytopes in $\mathcal{P}^4_8$ have at least
four obstruction vertices (which cannot be equilibrium
vertices).
Hence if both facets $P_1$ and $P_2$ belong to these 35 types, we
``lose'' at least four equilibria in each facet,
and hence there are at most
$40 - 4 - 4 = 32$
equilibria.
The two other combinatorial types are:
\rmitem{(i)}
the 4-cube, and
\rmitem{(ii)}
a type with 17 vertices that has only 3 obstruction
vertices, which we call the \textit{semi-cube}.

These two types play a prominent role in the subsequent investigations.
While the 4-cube is well-known, the semi-cube deserves
additional explanations. 
Three of its facets are 3-cubes. The graph of the semi-cube
is shown in Figure~\ref{fi:semi-cube}, where one can
identify two disjoint stable sets of size~7.
Each stable set contains two opposite corners from each of
the of the square faces. 
The semi-cube has also three disjoint
triangles, so the stable-set bound is 14 by the
disjoint-clique bound of Corollary~\ref{c:clique}.
In the classification of
\citet[p.~462]{gruenbaum-sreedharan-1967}, the semi-cube
appears as number~26, whereas in the data by
\citet{firsching-2017} the semi-cube appears as number~23. 

\ifpictures
\begin{figure}[t]
\[
\includegraphics[width=10cm]{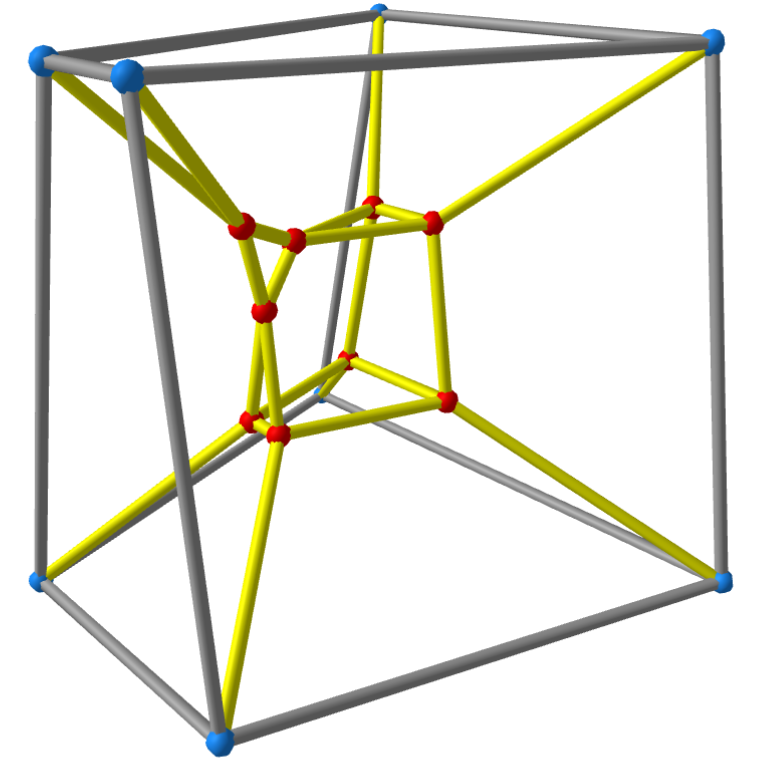}
\]
\caption{Schlegel diagram (see \cite{ziegler-1995})
and graph of the semi-cube.}
\label{fi:semi-cube}
\end{figure}
\fi

% \begin{lemma}
% \label{l:37}
The following is a list of the 37 combinatorial types of
polytopes in $\mathcal{P}^4_8$, with the numbering of
\citet{firsching-2017}.
Each pair $(V,b)$ states the number $V$ of vertices of the
polytope and its stable-set bound~$b$.
The number $V-b$ of obstruction vertices is at least 4
except for the underlined semi-cube number $23$ and cube
number $24$.
\begin{equation}
\arraycolsep.8em
\label{37}
 \begin{array}{lllll}
 1: (17,12) & 2: (18,12) & 3: (18,14) & 4: (18,12) & 5: (19,14) \\
 6: (17,12) & 7: (19,14) & 8: (16,10) & 9: (16,10) & 10: (16,12) \\
 11: (17,12) & 12: (19,14) & 13: (18,12) & 14:(16,10) & 15:(15,10) \\
 16: (20,14) & 17: (18,14) & 18:(19,14) & 19:(17,12) & 20:(20,16) \\
 21: (20,16) & 22: (18,12) & \underline{23:(17,14)} & \underline{24:(16,16)} 
 & 25: (17,12) \\
 26: (17,12) & 27:(17,12) & 28:(15,10) & 29:(16,10) & 30:(16,12) \\
 31: (15,10) & 32:(15,10) & 33:(15,8) & 34:(16,12) & 35:(14,8) \\
 36:(14,8) & 37:(14,8)
 \end{array}
\end{equation}
% \end{lemma}

We also have to consider disjoint facets $P_1$ and $P_2$ of
$P$ that as 4-polytopes have fewer than eight facets.
There are only five combinatorial types of polytopes in $\mathcal{P}_7^4$.
Similar to the pairs $(V,b)$ in (\ref{37}), their numbers $V$
of vertices and stable set bounds~$b$ are (via a computer
calculation, using data from \citet{firsching-2017}) 
\begin{equation}
\label{4-7}
(14,10), \quad
(13,8), \quad
(12,8), \quad
(12,8), \quad
(11,6) \, ,
\end{equation}
so each of them has also at least four obstruction vertices.
A polytope in $\mathcal{P}_6^4$ has at most nine vertices
and there are two possible types, with $(V,b)$ given by $(9,6)$ and
$(8,4)$. 
The only polytope in $\mathcal{P}_5^4$ is the 4-dimensional
simplex with $(V,b)=(5,2)$.

Hence, % assuming the correctness of the calculations, 
only the following polytopes $P$ are candidates for
obtaining more than 32 equilibria:

\rmitem{(i)}
$P_1$ or $P_2$ is a 4-cube.

\rmitem{(ii)}
$P_1$ or $P_2$ is a semi-cube.

\rmitem{(iii)}
$P_1$ or $P_2$ is a 4-polytope with only six facets and
$(V,b)=(9,6)$, or a simplex.

\noindent
In every other case the polytope $P$ has at least seven
obstruction vertices and therefore no more than 32
equilibrium vertices.

The following theorem is useful for understanding where 
equilibrium vertices in $Q$ can be found, which will be
applied to facets $Q_i$ of $Q$ using the facet-stable-set
bound of Theorem~\ref{t:facet}.
\def\Q{Q_i}

\begin{theorem}
\label{t:4-9}
Computer calculations show the following for 
every polytope $\Q$ in $\mathcal{P}^4_9$.
\rmitem{(a)}
The stable-set bound of $\Q$ is at most 20.
\rmitem{(b)}
$\Q$ has at least three obstruction vertices.
\rmitem{(c)}
If $\Q$ has stable-set bound 20, then $\Q$ does not have a 3-cube as a facet.
\rmitem{(d)}
If $\Q$ has stable-set bound 20, then $\Q$ 
has at least five obstruction vertices.
\end{theorem}

\myproof
The computer calculations are based on the list of the 1,142
combinatorial types of polytopes in $\mathcal{P}^4_9$ of
\citet{firsching-2017}.
The stable-set bound is computed using
Lemma~\ref{l:ilp}.
For (a), there are 39 types where the stable-set bound is~20.
The calculations are documented in \citet{stable-data}.
\endproof

We will use these properties to prove Theorem \ref{t:5x5}.
Case (iii) above is ruled out by the following lemma,
because $P_1$ or $P_2$ do not have enough vertices.

\begin{lemma}
\label{l:ruleout3}
Suppose $P$ is non-neighborly and has at least one facet
$P_i$ with a stable-set bound of less than 14.
Then the game has at most 32 equilibria.
\end{lemma}

\myproof
Let $P_i$ have label~$i$, and let $Q_i$ be the facet of $Q$
with label~$i$.
Every equilibrium $(x,y)$ needs to have label~$i$,
where by assumption there are fewer than 14 equilibrium
vertices~$x$.
In order to obtain 34 equilibria in total, the remaining
equilibria must have more than 20 equilibrium vertices $y$
in~$Q_i$.
By Theorem~\ref{t:4-9}(a), such a polytope $Q_i$ does
not exist in $\mathcal{P}^4_9$.
The same applies if the 4-polytope $Q_i$ has eight or fewer
facets, because then it has at most 20 vertices.
\endproof

We next treat case (ii) above.

\begin{lemma}
\label{l:semi}
Let $P$ and $Q$ be non-neighborly, with disjoint facets $P_1$ and
$P_2$ of $P$, and let $P_1$ be a semi-cube.
Then the game has at most 32 equilibria. 
\end{lemma}

\myproof
Assume that the game has 34 or more equilibria (their number
is even by Theorem~\ref{t:even}), which will
lead to a contradiction.
The stable-set set bound for $P_1$ is 14 by (\ref{37}),
and $P_1$ needs to have 14 equilibrium vertices, by
Lemma~\ref{l:ruleout3}.
The remaining 20 equilibrium vertices with label~1 belong to~$Q_1$,
which is in $\mathcal P_9^4$ by Theorem~\ref{t:4-9}(a) (by
(\ref{37}), no polytope in $\mathcal P_8^4$ has 20
equilibrium vertices).
Hence, $Q_1$ intersects with every other facet of $Q$, where
the resulting ridge (a three-dimensional face both of~$Q$
and of any facet of~$Q$) is never a 3-cube by
Theorem~\ref{t:4-9}(c).
Hence, none of the facets of~$Q$ is a 4-cube.
In order to obtain more than 32 equilibria, the 
two disjoint facets of $Q$ are therefore semi-cubes, because
combining a semi-cube with any of the other combinatorial
types in (\ref{37}) would give at least $3+4$ obstruction
vertices.
This argument with the polytopes exchanged implies that
$P_2$ is also a semi-cube.
Hence, also $Q_2\in\mathcal P_9^4$.

By Theorem~\ref{t:4-9}(d),
both $Q_1$ and $Q_2$ have 20 equilibrium vertices and at least 
five obstruction vertices.
By Theorem~\ref{t:4-9}(d) and (b) and~(\ref{37}),
each of the other eight facets of
$Q$ has at least three obstruction vertices.
Counted by multiplicity over all ten facets (each vertex
belongs to five facets), the number of obstruction
vertices is therefore at least
$(2\cdot 5+8\cdot 3)/5=34/5>6$.
Hence, $Q$ has at least seven obstruction vertices,
but at most 40 vertices, and therefore at most 32
equilibria.
\endproof

Due to the symmetry of the role of the two polytopes, case
(ii) and the previous lemma cover the case that at least one
of the disjoint facet pairs of $P$ or $Q$ contains a semi-cube.
The remaining case (i) is covered by the following lemma.

\begin{lemma}
\label{l:16facets}
Let $P$ and $Q$ be non-neighborly and assume that one of the
disjoint facets $P_1,P_2$ is 4-cube, and
that one facet of $Q$ is a 4-cube.
Then the game has at most 32 equilibria.
\end{lemma}

\myproof
Suppose $P_1$ is a 4-cube.
Assume that the game has 34 or more equilibria, which 
will lead to a contradiction.
If $P_1$ had 15 or fewer equilibrium vertices, then the
remaining equilibria have label~1 with their 19 equilibrium
vertices in $Q_1$, which needs to have a stable-set bound of
20 but then has no 3-cube as a facet by Theorem~\ref{t:4-9}(c).
However, then $Q$ has no 3-cube as a ridge, which contradicts $Q$
having a 4-cube as a facet.
Hence, \textit{all} 16 vertices of $P_1$ are equilibrium
vertices.

The 4-cube $P_1$ has a special combinatorial structure, which
is isomorphic to that of the unit 4-cube
$W=\{z\in\reals^4\mid 0\le z_i\le 1$ for $i\in[4]\}$.
Each of the eight inequalities of $W$ corresponds to a facet of
$P_1$ with a label in $\{3,\ldots,10\}$.
These facets of $P_1$ come in four pairs with labels $g_i,h_i$
that correspond to the binding inequalities $z_i=0$ and
$z_i=1$ in~$W$ for $i\in[4]$.
Every vertex $v$ of $P_1$ has exactly one label $g_i$ or
$h_i$ for $i\in[4]$, and has a \textit{complementary} vertex
(diagonally opposite in~$W$) defined by exactly the other
labels from each pair.
Then the {partner} of $v$ in $Q$ has these four
complementary labels, as well as label~2, so all 
the partners of $P_1$ are in~$Q_2$.
The complementary vertices in $P_1$ themselves have also
their partners in~$Q_2$.
The resulting labels show that the partners of all the
vertices of $P_1$ represent a 4-cube in~$Q_2$.
In particular, all of them have four neighbors (via their
edges) in $Q_2$.
Because $Q_2$ is a simple 4-polytope, it does not have any
further vertices because they could not be connected to
these 16 vertices.
Hence, $Q_2$ is itself a 4-cube and has exactly 16
equilibrium vertices.

Every equilibrium $(x,y)$ has to have label~2.
If $y$ has label~2 then $y\in Q_2$.
The other equilibria $(x,y)$ require that $x\in P_2$, where
$P_2\in\mathcal P_8^4$, but $P_2$ has at most 16 equilibrium
vertices by~(\ref{37}).
\endproof

Lemmas~\ref{l:ruleout3},~\ref{l:semi}, and~\ref{l:16facets}
cover the cases (iii), (ii), (i) of the pairs $P_1,P_2$ of
disjoint facets of $P$ that together have fewer than seven
obstruction vertices.
This completes the proof of Theorem~\ref{t:5x5}.

\subsection*{Acknowledgments}

The third author is indebted to Vissarion Fisikopoulous for
our first attempt at this problem in 2015.
That attempt provided an independent proof of
Theorem~\ref{t:neighb} but was incomplete for proving
Theorem~\ref{t:5x5}, because we could
not resolve the case of vertices that do not belong to
either of the disjoint facets in a non-neighborly polytope
in $\mathcal P_{10}^5$.
We also thank Moritz Firsching for helpful communication on his
classification of 5-polytopes.

This work was supported by the DFG Priority Program
``Combinatorial Synergies'' (grant no.\ 539847176).

% \small
% \bibsep1ex plus.1ex minus.05ex
% \bibliography{bib-nash-combinatorics}
% \bibliographystyle{book}
% 
% \end{document}

\small
\bibsep.9ex plus.1ex minus.05ex
\bibliography{bib-nash-combinatorics}
\bibliographystyle{book}

\end{document}